# Deriving Autism spectrum disorder functional networks from rs-fMRI data using group ICA and dictionary learning

Xin Yang[1], Ning Zhang[2] and Donglin Wang[3]

[1]Department of Computer Science, Middle Tennessee State University,
Murfreesboro, TN, USA
[2]Department of Computer Information Sciences, St. Ambrose University,
Davenport, USA
[3]Department of Mathematical Sciences, Middle Tennessee State University,
Murfreesboro, TN, USA

## Abstract

*The objective of this study is to derive functional networks for the autism spectrum disorder (ASD) population using the group ICA and dictionary learning model together and to classify ASD and typically developing (TD) participants using the functional connectivity calculated from the derived functional networks. In our experiments, the ASD functional networks were derived from resting-state functional magnetic resonance imaging (rs-fMRI) data. We downloaded a total of 120 training samples, including 58 ASD and 62 TD participants, which were obtained from the public repository: Autism Brain Imaging Data Exchange I (ABIDE I). Our methodology and results have five main parts. First, we utilize a group ICA model to extract functional networks from the ASD group and rank the top 20 regions of interest (ROIs). Second, we utilize a dictionary learning model to extract functional networks from the ASD group and rank the top 20 ROIs. Third, we merged the 40 selected ROIs from the two models together as the ASD functional networks. Fourth, we generate three corresponding masks based on the 20 selected ROIs from group ICA, the 20 ROIs selected from dictionary learning, and the 40 combined ROIs selected from both. Finally, we extract ROIs for all training samples using the above three masks, and the calculated functional connectivity was used as features for ASD and TD classification. The classification results showed that the functional networks derived from ICA and dictionary learning together outperform those derived from a single ICA model or a single dictionary learning model.*

## Keywords

*Functional connectivity, rs-fMRI, autism spectrum disorder (ASD), group ICA, Dictionary Learning.*

## 1. Introduction

The human brain is very mysterious to humans since it is the most complex organ known in the world. Although attempts have been made for centuries to study and unravel the mystery of the human brain, our understanding of it is still limited. Therefore, a deep understanding of the brain is essential if humans are to unravel the relationship between brain function in neurological function-related disease. Functional magnetic resonance imaging (fMRI) is a technique that uses magnetic resonance imaging to detect brain activity by calculating the fluctuations in local blood oxygenation level, which provides a means of understanding how spatially distributed brain regions interact and work together to create neurological function.

Prior to the development of functional neuroimaging methods, a series of developments and advances were experienced. In the 1970s, Allan M. Cormack and Godfrey N. Hounsfield invented the computer-assisted tomography imaging technique. This technique allows the acquisition of higher resolution images of brain structures. Soon after, the invention of radioligands led to two new neuroimaging techniques: single photon emission computed tomography (SPECT) and positron emission tomography (PET). MRI is a relatively new medical imaging technique. In the early years of the 21st century, developments in neuroimaging began to allow functional neuroimaging techniques. Compared to ionizing radiation methods such as computed tomography (CT)and positron emission tomography (PET), fMRI offers a completely safe and non-invasive method of imaging brain activity with reasonable spatial and temporal resolution. With the rapid development of functional magnetic resonance imaging (fMRI), modern cognitive neuroscientists now have an imaging tool that overcomes the limitations of earlier neurocognition studies. Since then, the field of fMRI has led to remarkable progress in biomedical research [1].

Researchers were most initially interested in fMRI for the brain's response to external mental stimulation. Thus, most of the initial studies focused on the response to external task-evoked activities [2]. In 1995, Biswal et al [3,4] found that the primary motor cortex's left and right hemispheric regions were not in a silent state at rest. They were the first to hypothesize that the functional connectivity exhibited in the motor cortex is a general phenomenon and not due to external stimulus events. Their discoveries suggested that resting-state fMRI can also provide meaningful information on neurological function even in the absence of external events stimulation. Since then, there has been an explosion of subsequent studies of brain function using resting-state fMRI (rs-fMRI) data [5,6,7]. Scientists have come to recognize the usefulness of studying whether patterns identified in resting-state fMRI data exhibit the same characteristics under different conditions. Nowadays, rs-fMRI has become an essential technique for analyzing neurological disorders such as Alzheimer's disease, autism spectrum disease, etc.

Autism spectrum disorder (ASD) is a neurologically impaired brain disorder in which individuals have impaired development of social interaction and communication skills [8]. According to Centers for Disease Control (CDC), to date, approximately one in every 54 children has been diagnosed with an autism spectrum disorder in the United States. Research in autism spectrum disorder is imperative as more and more families around the world are affected by the disorder. The diagnosis of autism spectrum disorder is challenging and complex. The clinical approach to diagnosis is generally based on a comprehensive behavioral assessment by a child psychiatrist or psychologist, which includes observation of the child's behavior, speech and language, hearing, vision, motor function, etc. [8] The clinical approach has shed light on many aspects of autism spectrum disorder in behavior perspective.

Although brain neurologists and neuroscientists have suggested many causes, including genetic and environmental factors, the exact etiology of autism spectrum disorder remains unclear. In addition to the behavioral assessment, recent advances in neuroimaging techniques have prompted the possibility of interpreting the connectivity between behavioral disorders and neurological function from fMRI data. An increasing number of research demonstrates that social and communicative deficits are associated with the function and connectivity of cortical networks.

The theory of cortical under-connectivity has been proposed as an explanatory model for ASD, suggesting that abnormal functional connectivity between brain regions may contribute to poor performance on cognitive and social tasks in people with ASD [9, 10,11,12].

To explore the difference in brain connectivity between ASD and TD groups. In this study, we aim to compare functional connectivity in ASD groups and TD groups. We combined group independent component analysis (ICA) and dictionary learning together to identify functional networks and investigate their connectivity. ICA is a data-driven method that separates a multivariate signal into additive subcomponents [13]. ICA attempts to decompose a multivariate signal into a linear combinations of independent non-Gaussian signals. When ICA is applied to fMRI data, the 4D fMRI time series signal are typically modeled as linear combinations of unknown spatially independent activity patterns. The 4D fMRI time series signals are decomposed into spatially independent components (ICs), but temporally coherent networks. Spatial ICA has been applied to resting-state fMRI of anesthetized child patients by Kiviniemi. By analyzing the statistical characteristics of the observed data samples and minimizing the mutual information between the observed signals, ICA can separate out different source signals [14]. The problem arises, however, that ICA is required to comply with the orthogonality constraints on the data representation subspace, which leads to the fact that the maximum number of causes is often limited to the signal dimension. In response to this situation, this problem has triggered the emergence of a new promising research area, namely dictionary learning. The focus of Dictionary learning is to construct a dictionary of atoms or subspaces that provides efficient representations for the observed samples [14]. Dictionary learning has the potential to derive the priori unknown statistics for sparse signals. It has been successfully applied in the field of medical imaging, such as electroencephalogram (EEG), magnetic resonance imaging (MRI), and functional MRI (fMRI).

The objective of this study is to derive functional networks from autism spectrum disorder (ASD) groups using group ICA and Dictionary Learning and use the functional connectivity calculated from the derived functional networks to classify ASD and TD groups. In our experiments, the ASD functional networks were derived from resting-state functional magnetic resonance imaging (rs-fMRI) data. We downloaded a total of 120 training samples including 58 ASD and 62 TD participants, which were obtained from the Autism Brain Imaging Data Exchange I (ABIDE I) repository. The functional connectivity matrix was calculated from the derived functional networks, which have been applied as the classification features. Our methodology and results have five main parts. First, we utilize the group ICA model to extract ASD functional networks and rank the top 20 ROIs. Second, we utilize a dictionary learning model to extract ASD functional networks and rank the top 20 ROIs. Third, we merged the 40 selected ROIs from the two models together as the ASD functional networks. Fourth, we generate three corresponding masks based on the 20 selected ROIs from group ICA, the 20 ROIs selected from dictionary learning, and the 40 combined ROIs selected from both. Finally, we extract ROIs for all training sample using the above three masks, and the calculated functional connectivity was used to classify ASD and TD participants. The classification results showed that the functional networks derived from ICA and dictionary learning together outperform those derived from a single ICA model or a single dictionary learning model.

## 2. METHODS

### 2.1. Datasets

A total of 120 participants, 58 with a diagnosis of ASD and 62 TD participants, were included in this study. All data were obtained from the public repository: Autism Brain Imaging Exchange I (ABIDE I). The ABIDE I represents the first ABIDE initiative. The ABIDE I datasets consists of structural MRI and resting-state fMRI data, and the corresponding phenotypic information. The fMRI data from ABIDE I were pre-processed using Configurable Pipeline for the Analysis of Connectomes (CPAC). CPAC is an open-source pipeline to pre-process resting-state fMRI data.

In these 120 subjects, there are 58 ASD and 62 TD subjects, of which 13 females and 107 males. The summary information of the selected 120 subjects is displayed in Table I. Table I contains a summary of phenotypic information for ASD and TD, such as sex, age, and experimental site name.

Table 1: ABIDE data phenotypical information summary

| Site | Count | | Count | | Total | Age Range |
|------|-----|----|-----|---|-------|-----------|
|      | ASD | TD | M   | F |       |           |
| OHSU | 12  | 13 | 25  | 0 | 25    | 8~15      |
| OLIN | 14  | 14 | 23  | 5 | 28    | 10~24     |
| PITT | 24  | 26 | 43  | 7 | 50    | 9~35      |
| SDSU | 8   | 9  | 16  | 1 | 17    | 12~17     |
| TOTAL| 58  | 62 | 107 | 13| 120   | 8~35      |

In this study, we aim to combine group ICA and dictionary learning together to identify functional networks and investigate ASD and TD classification using functional connectivity calculated from the derived functional networks.

## 2.2. Spatial Independent Component Analysis on single-subject fMRI data

ICA is a blind source separation method for separating data into underlying informational components [14]. ICA assumes that the observed data samples can be decomposed into linear combinations of unknown underlying signals and that the data can be reconstructed based on statistical independence. Accordingly, ICA separates signal mixtures into statistically independent signals. ICA has been applied to different domains such as speech processing, neuroimaging (fMRI, EEG), telecommunications, and stock market prediction [15].

The ICA model is a statistical model with linear combinations of mixed signals. When we use the data matrix $X$ to represent the observed data, the model is generally expressed in matrix form as following:

$$X = AS$$

where $X = (X_1, \ldots, X_m)^T$ is the observed data matrix with dimension $T \times M$, $A = (a_{ij})$ is the unknown mixing matrix of size $T \times K$, and $S = (S_1, \ldots, S_m)^T$ is the $m$ unknown source signals need to be recovered. In total, there are $K$ sources. In the source matrix $S$, each row $s_k^T$ represents an independent component. In the mixing matrix $A$, each column $a_k$ represents its corresponding weights. This can be written in the form of linear weighted sums: $x_t = a_{t1}s_1 + \cdots + a_{tk}s_k$.

It's well-known that ICA algorithms are generally widely used for time series signals. However, it must be emphasized that the fMRI signals are time series of the spatial volumes, and the different activation patterns derived from the fMRI signals are also spatial-oriented. Therefore, spatial ICA is more suitable for fMRI data analysis [16]. The spatial ICA model is illustrated in Figure 1:

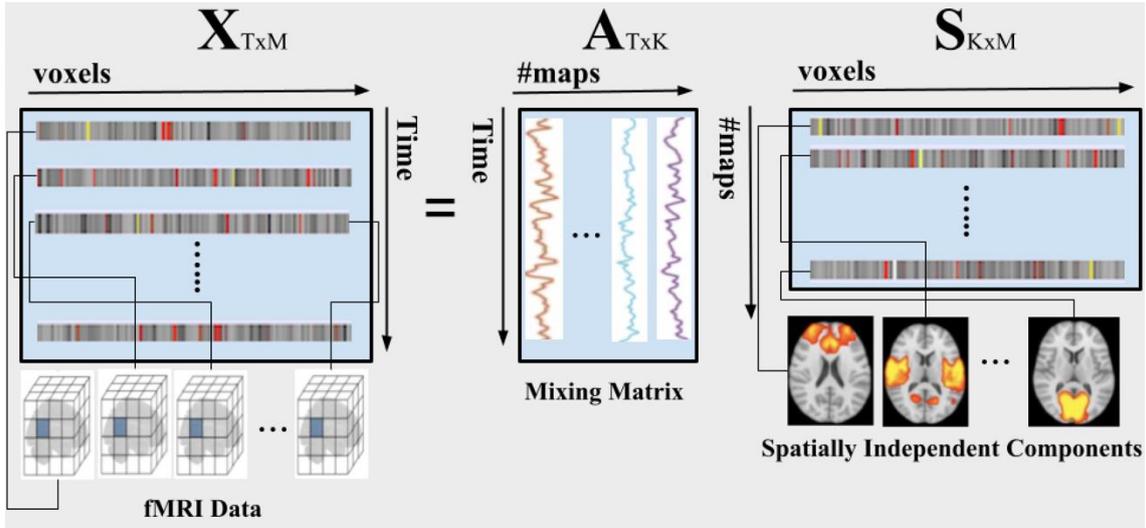

Figure 1: Spatial ICA for single fMRI data

From Figure 1 we can see that each row in the data matrix $X$ represents a volume vector, and each row in the source matrix $S$ represents an independent spatial pattern. In the mixing matrix $A$, each column represents the activation time series. The $T \times M$ data matrix $X$ represents the observed fMRI data. Here, $M$ is the total number of voxels in a subject's brain and $T$ is the number of fMRI time series points.

## 2.3. group ICA on multi-subject fMRI data

In the context of the prevalence of group analysis, a large number of studies on rs-fMRI have shown that the correlation patterns found in the BOLD signals are highly reproducible across populations [20]. Meanwhile, some of the network patterns derived from rs-fMRI by the ICA model are consistent at the group level. However, the ICA model can be sensitive to data variations, even just mild variations. Based on these factors, a direct comparison of patterns estimated from different individual subject is not meaningful. Instead, it would be more meaningful and reasonable to analyze group-level patterns specifically for each subject. For the group-level extraction of ICA patterns, researchers have adopted different strategies. The data volumes from Individual subjects can be concatenated together in a time series, and then the ICA model can be applied to the group data [22]. Beckmann and Smith [23] proposed a novel model that refers to tensorial extension of ICA, which will estimate the patterns across subjects in the same time course. However, these methods can not directly detect the difference between groups in terms of individual ICs. In order to be able to represent the variability due to individual differences between subjects, Varoquaux etc. proposed an updated group model, called CanICA, to extract group-level IC components. The advantage of their proposed model is that by using generalized canonical correlation analysis (CCA) it can identify a subspace of reproducible components across subjects [20].

The observed time series fMRI data for each subject $Y_s$ can be consist of a set of independent spatial patterns $P_s$ with observation noise. For each subject $Y_s$, it takes the form:

$$Y_s = W_s P_s + E_s$$

, where $W_s$ is a loading matrix, and $E_s$ is the observation noise. Each subject $Y_s$ activity can be described by subject-specific spatial patterns $P_s$, which are a combination of the group-level patterns $B$ and additional subject-variability, the $P_s$ takes the form:

$$P_s = A_s B + R_s$$

The group form can be written with vertically concatenated matrices: $P = \{P_1, P_2, \dots, P_s\}$, $R = \{R_1, R_2, \dots, R_s\}$, and $A = \{A_1, A_2, \dots, A_s\}$, $s = 1, \dots, N$. The canICA model is illustrated in Figure 2. The following Figure describes the steps starting from individual fMRI subjects, to obtain the group-level independent components.

$$P = AB + R$$

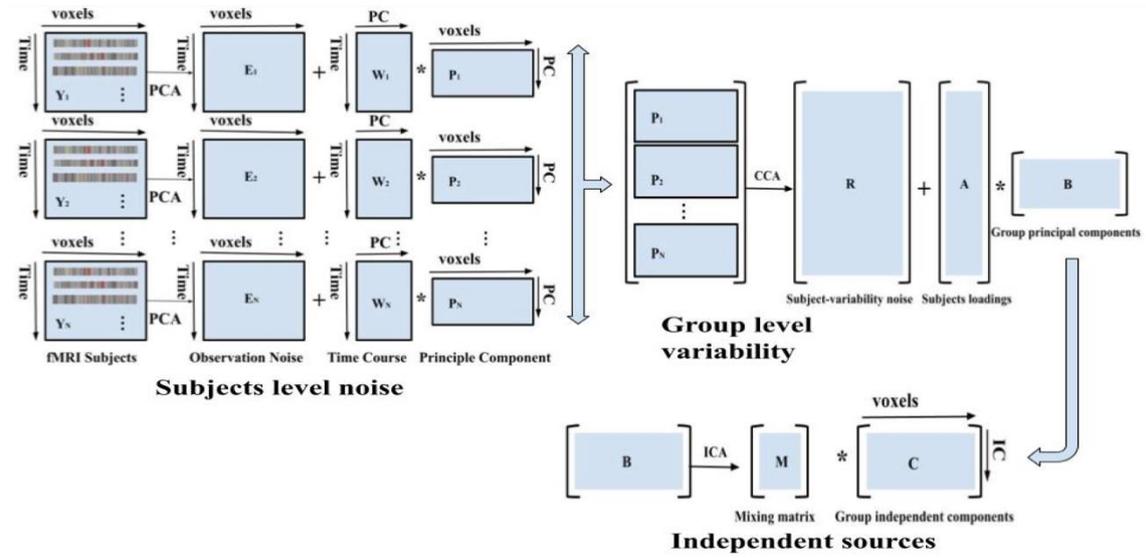

Fig 2: The steps starting from individual fMRI subjects, to obtain group-level independent components.

## 2.4. Dictionary Learning

Recently, dictionary learning and sparse representation have been shown to be efficient in the machine learning and pattern recognition fields [17, 18, 19, 21]. The purpose of the sparse representation is to learn a set of basis vectors and to represent the original signals using a linear combination of these basis vectors [17]. At the same time, a variety of neuroscience studies have reported the presence of sparse response in the brain neural activity [18]. The sparse response of neural activity in the brain coincides with the intrinsic nature of sparse representation methods, suggesting that sparse representation may be a possible solution to brain activity detection.

Figure 3 summarizes the framework of investigating functional networks via dictionary learning and sparse representation. Given the rs-fMRI signal matrix $S$, where $M$ is the total number of voxels in a subject's brain and $T$ is the total number of fMRI time series points [19], each rs-fMRI signal in $S$ is modelled as a linear combination of the learned basis dictionary $D$, the model is expressed in matrix form as following:

$$S = DA$$

, where **A** is holding the coefficient matrix for sparse representation. Specifically, the signal of each dictionary atom in **D** represents the functional activity of a specific brain network, while the vector in the coefficient matrix **A** represents the spatial distribution of the corresponding brain network. Finally, we identify functional network components by performing the components of interests in the learned dictionary **D**.

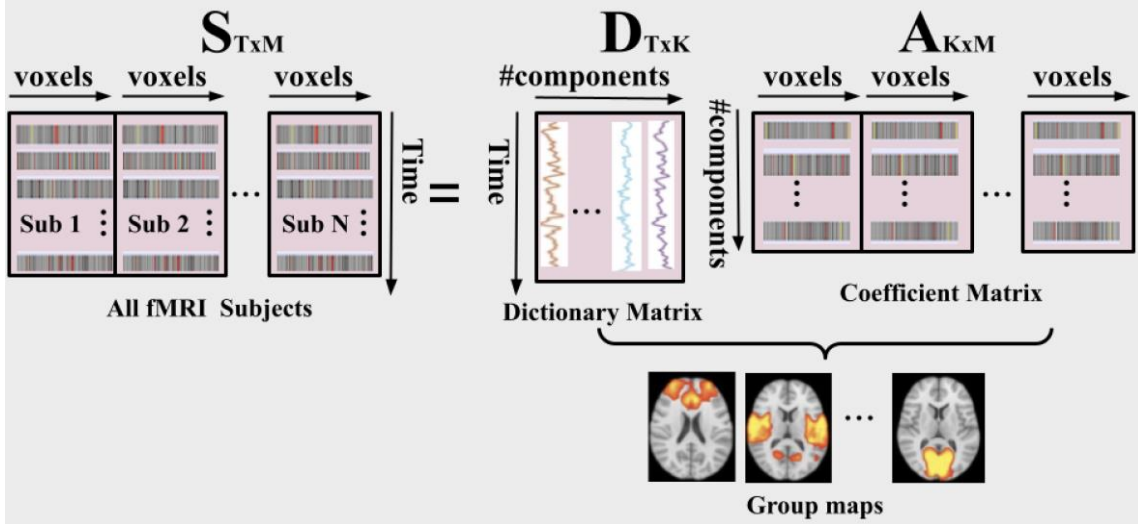

Figure 3: The framework for group-level analysis using Dictionary Learning

## 3. EXPERIMENTAL RESULTS

Our method and results have five main parts. First, we extracted ASD functional networks using the grouped ICA model and ranked the top 20 ROIs. Second, we extracted ASD functional networks using a dictionary learning model and ranked the top 20 ROIs. Third, we merged the 40 selected ROIs from the two models together as ASD functional networks. Fourth, we generate three corresponding masks based on the 20 selected ROIs from group ICA, the 20 ROIs selected from dictionary learning, and the 40 combined ROIs selected from both. Finally, we extract ROIs for all training sample using the above three masks, and the calculated functional connectivity was used to classify ASD and TD participants. The experimental codes for this paper are available at GitHub: https://github.com/XinYangMTSU/BIGML

### 3.1. Group Independent Component Analysis (group ICA)

First, the pre-processed 4D fMRI time series data from the ASD group were analyzed using the group ICA to identify spatially independent and temporally coherent networks. Figure 4 shows all 20 independent components (ROIs) derived from group ICA. Figure 5 shows each independent component individually in axial view. Figure 6 shows the ranked importance of all extracted ICA components (ROIs), and the ranking is done using the explained variance. We generate a group ICA mask based on the selected 20 ROIs.

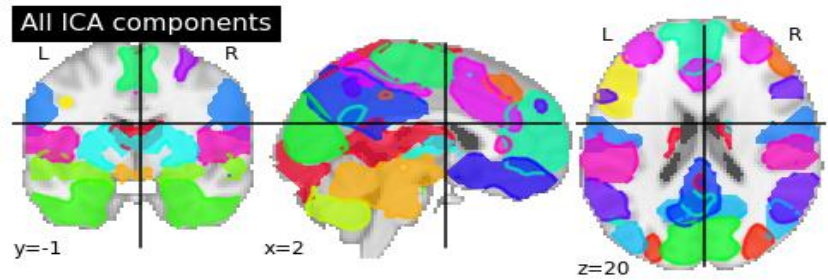

Figure 4: 20 independent components extract from group ICA

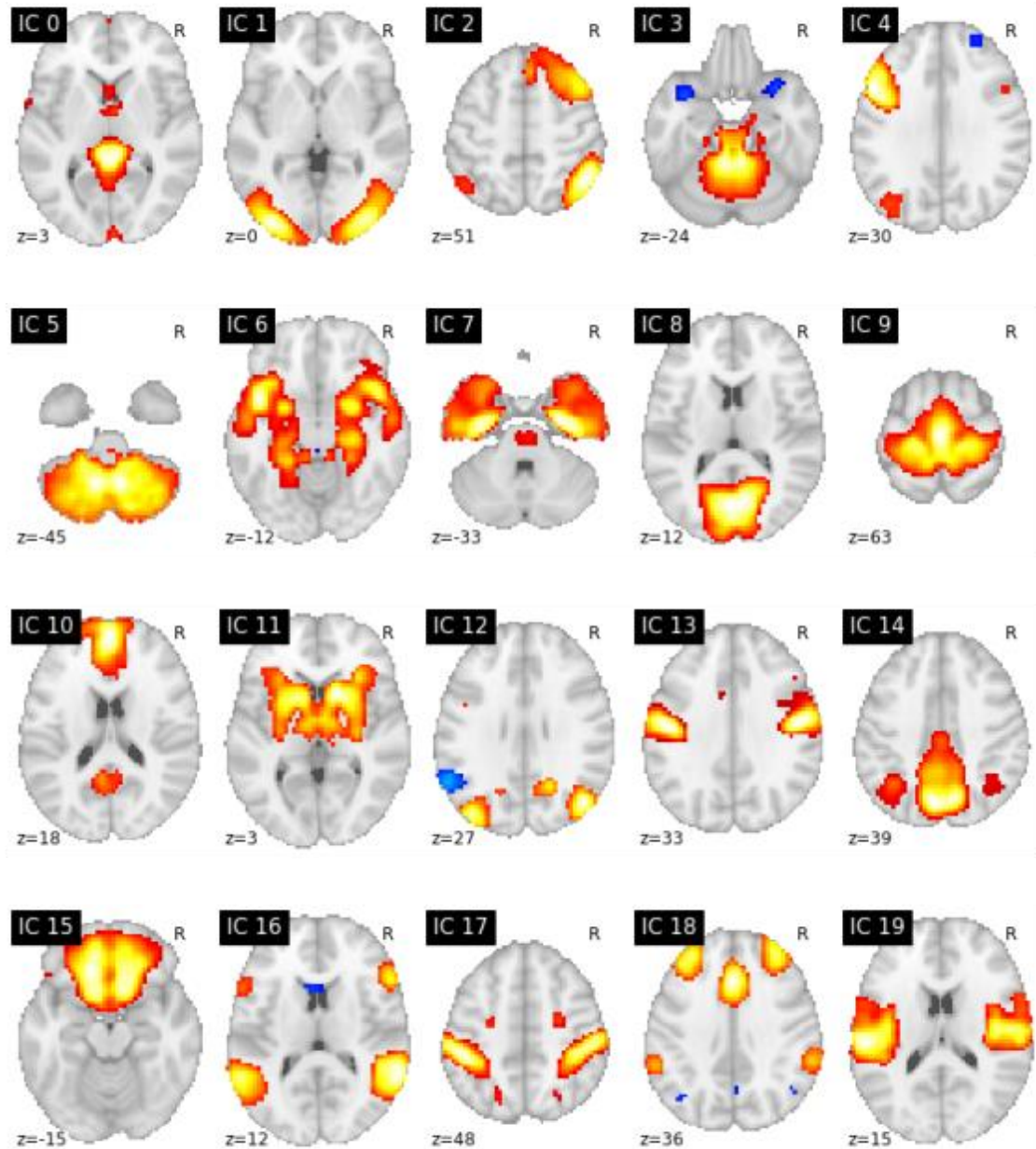

Figure 5: The 20 ICA maps derived using CanICA from rs-fMRI

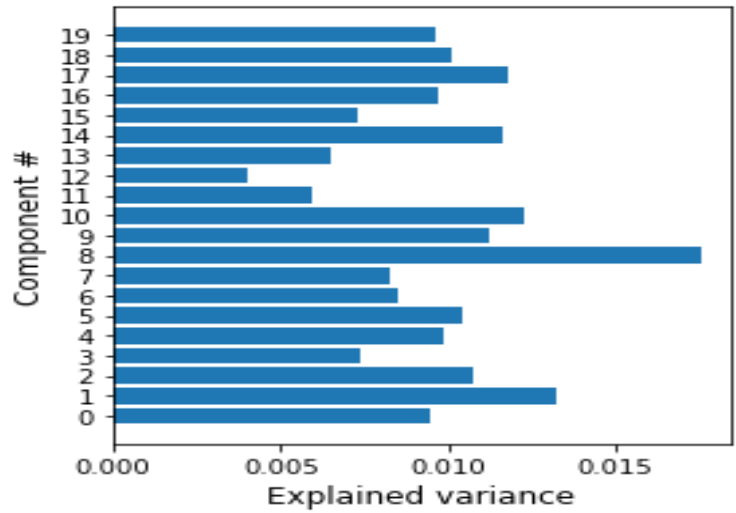

Figure 6: Explained variance of all 20 ICA components

## 3.2. Dictionary Learning

Second, the pre-processed 4D fMRI time series data from the ASD group were analyzed using the dictionary learning model to identify spatially independent and temporally coherent networks. Figure 7 shows all 20 independent components (ROIs) derived from dictionary learning. Figure 8 shows each independent component individually in axial view. Figure 9 shows the ranked importance of all extracted components (ROIs), and the ranking is done using the explained variance. We generate a dictionary learning mask based on the selected 20 ROIs.

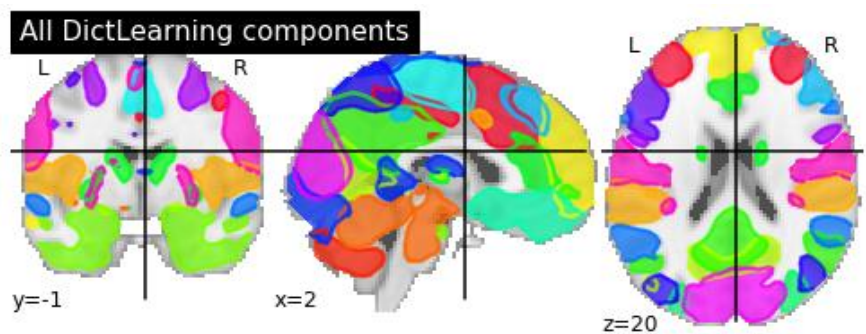

Figure 7: 20 independent components extract using dictionary learning from rs-fMRI

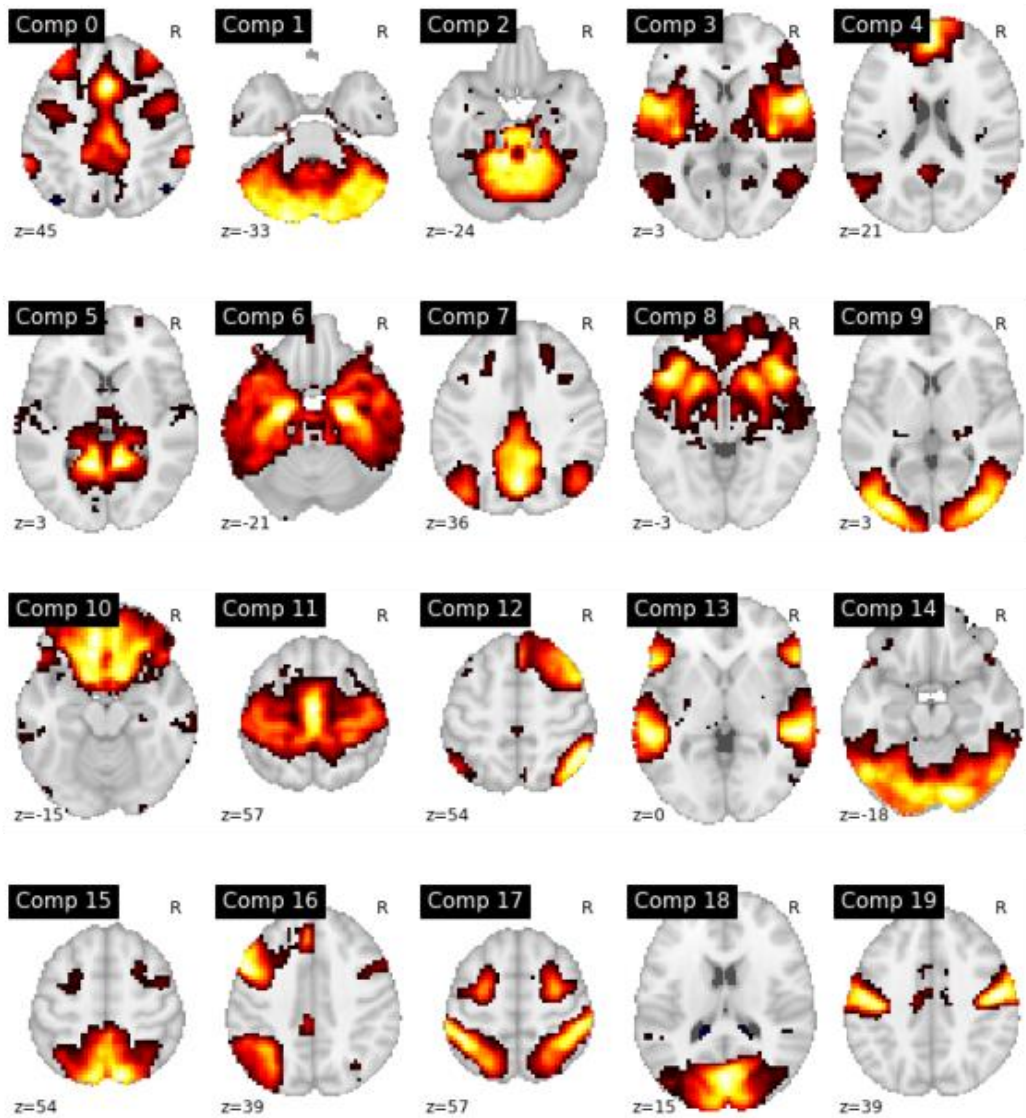

Figure 8: The 20 maps derived by dictionary learning from rs-fMRI

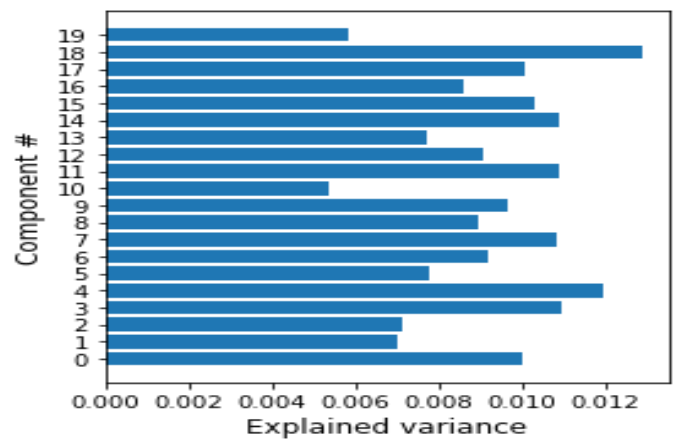

Figure 9: Explained variance of all 20 dictionary learning components

### 3.3. Group ICA and dictionary learning (ICA+Dict)

Third, we combined the 40 ROIs obtained from group ICA and dictionary learning together as the ASD functional networks. We generate an ICA+Dict mask based on the selected 40 ROIs.

We extracted all 40 ROIs from the rs-fMRI time series data and then calculated the functional connectivity among all ROIs. In this study, the functional connectivity calculated from the derived functional networks was used to classify ASD and TD subjects.

### 3.4. ASD and TD classification using functional connectivity

We generate three corresponding masks based on the 20 selected ROIs from group ICA, the 20 ROIs selected from dictionary learning, and the 40 combined ROIs selected from both. Based on the above three generated masks, we can extract corresponding ROIs for all training samples and calculate the pairwise functional connectivity. The calculated functional connectivity was used to classify ASD and TD participants.

In order to calculate the connectivity matrix of the ROIs, we can implement the Pearson correlation. Every functional connectivity feature in the connectivity matrix is represented by a Pearson correlation coefficient, which is used to measure the mutual relationship between two brain regions of interest. The threshold of Pearson's correlation coefficient ranges from -1 to +1. When the threshold value of the Pearson correlation coefficient approaches -1, this indicates an opposite association between the two brain regions; in contrast, if the Pearson coefficient value approaches 1, it indicates a high correlation between the two brain regions. It is worth noting that the Pearson correlation connectivity matrix is symmetric, which means that the corresponding upper and lower triangular values are the same. Therefore, we only need to use the upper triangular or lower triangular values in the correlation matrix as features for the ASD and TD classification. In addition to this, the main diagonal in the connectivity matrix should also be removed, as these values represent self-correlated regions.

For single group ICA and single dictionary learning, the ROI-based functional connectivity of the extracted 20 functional ROIs was calculated, resulting in 190 ($\frac{(400-20)}{2} = 190$) pair-wise connectivity features for each subject. This set of 190 features for each subject was used as features for ASD and TD classification.

For ICA+Dict, the ROI-based functional connectivity of the extracted 40 functional ROIs was calculated, resulting in 780 ($\frac{(1600-40)}{2} = 780$) pair-wise connectivity features for each subject. This set of 780 features for each subject was used as features for ASD and TD classification.

In our experiments, we use the Gaussian kernel support vector machine (kSVM) to classify ASD and TD. To evaluate the performance, we did five-fold cross-validation. Table 2 shows that the 5-fold cross-validation sensitivity, specificity and accuracy of ICA+Dict are better than that of single ICA or single dictionary learning. It is noteworthy that the sensitivity of ICA+Dict is improved by at least 3% compared to single ICA or single dictionary learning. For ASD studies, sensitivity is a more important metric than specificity. Therefore, overall, ICA+Dict outperforms either single ICA or single dictionary learning. Figure 10, 11 and 12 shows the five-fold cross-validation receiver operating characteristic (ROC) curve for group ICA, dictionary learning, and ICA+Dict respectively.

Table 2: Five-fold Cross-Validation Results

|  | Sensitivity | Specificity | Accuracy |
|---|---|---|---|
| kSVM-ICA | 55.61% | 58.08% | 56.67% |
| kSVM-Dict | 65.76% | 69.10% | 67.50% |
| kSVM-ICA+Dict | **69.24%** | **69.23%** | **69.17%** |

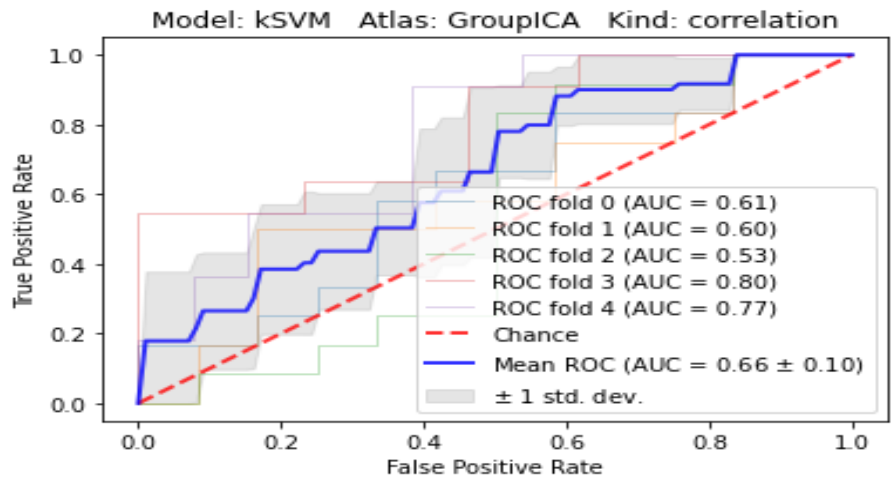

Figure 10: Five-fold cross validation ROC Curve for groupICA

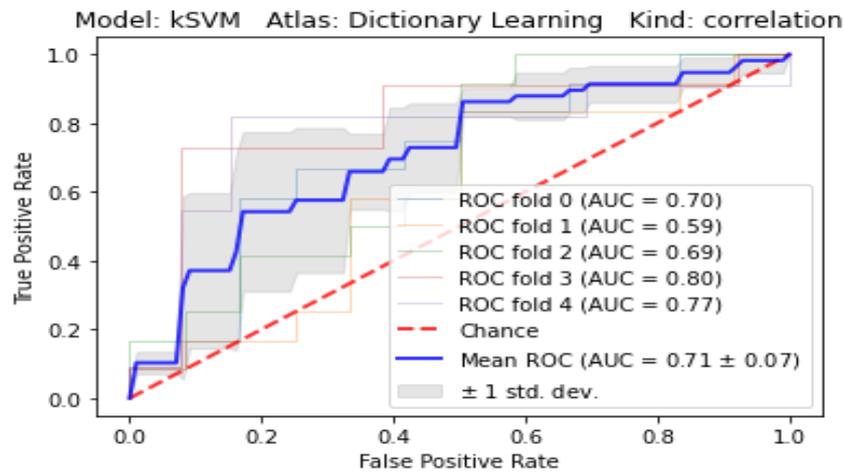

Figure 11: Five-fold cross validation ROC Curve for Dictionary Learning

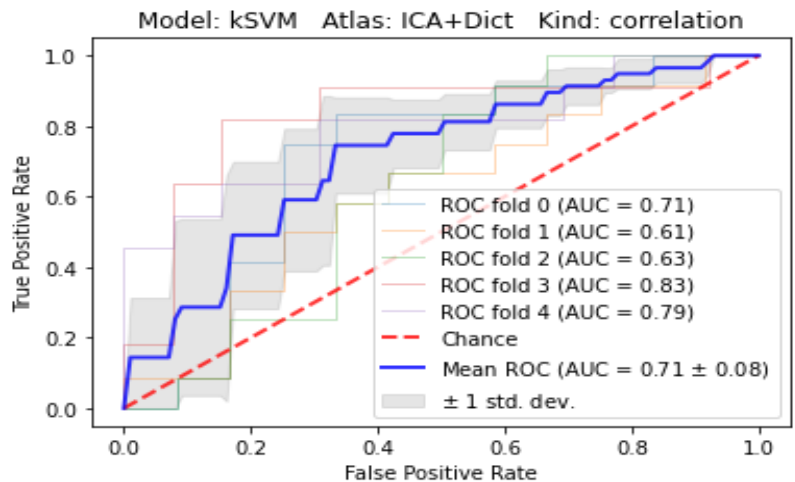

Figure 12: Five-fold cross validation ROC Curve fro group ICA and Dictionary Learning

## 4. CONCLUSIONS

In this study, we used group ICA and dictionary learning model together to derive functional networks for the autism spectrum disorder (ASD) population, and the functional connectivity calculated from the derived functional networks is used as the feature to classify ASD and TD participants. We combined the 40 ROIs obtained from group ICA and dictionary learning together as the ASD functional networks. The ROI-based functional connectivity of the extracted 40 functional ROIs was calculated, resulting in 780 pair-wise connectivity features for each subject. This set of 780 features for each subject was used as features for ASD and TD classification. The 5-fold cross-validation results showed that the functional networks derived from ICA and dictionary learning together obtain higher sensitivity, specificity and accuracy than a single ICA model or single dictionary learning model. Our results demonstrate that the improved data-driven algorithm can extract ROIs containing important information to improve the accuracy of ASD and TD classification. This also shows that data-driven algorithm is still a promising research direction for the field of ASD research.

In this experiment, our sample size of 120 is still relatively small. To get more accurate and stable results, we need to collect more samples to conduct the experiment. Nowadays, deep learning has been widely used in various fields. Nonetheless, supervised deep learning requires a training sample set of sufficient size. Our further research in ASD classification will require the need to collect a large amount of training data to achieve considerable breakthroughs in sensitivity, accuracy, and specificity.

## REFERENCES

[1] Poldrack, R.A., Mumford, J.A. and Nichols, T.E., 2011. Handbook of functional MRI data analysis. Cambridge University Press.

[2] Poldrack RA, Mumford JA, Nichols TE. Handbook of functional MRI data analysis. Cambridge University Press; 2011 Aug 22.

[3] Biswal, B., ZerrinYetkin, F., Haughton, V.M. and Hyde, J.S., 1995. Functional connectivity in the motor cortex of resting human brain using echo-planar MRI. Magnetic resonance in medicine, 34(4), pp.537-541.

[4] Biswal BB, Van Kylen J, Hyde JS. Simultaneous assessment of flow and BOLD signals in resting-state functional connectivity maps. NMR in Biomedicine. 1997 Jun-Aug;10(4-5):165-170.

[5]   Damoiseaux, Jessica &Rombouts, Serge &Barkhof, Frederik &Scheltens, Ph & Stam, C.J. & Smith, S.M. & Beckmann, Christian. (2006). Consistent resting-state networks. Proceedings of the National Academy of Sciences of the United States of America. 103. 13848-53. 10.1073/pnas.0601417103.

[6]   Greicius MD, Srivastava G, Reiss AL, Menon V. Default-mode network activity distinguishes Alzheimer's disease from healthy aging: evidence from functional MRI. Proc Natl Acad Sci U S A. 2004 Mar 30;101(13):4637-42. doi: 10.1073/pnas.0308627101. Epub 2004 Mar 15. PMID: 15070770; PMCID: PMC384799.

[7]   Al-Zubaidi, A., Mertins, A., Heldmann, M., Jauch-Chara, K. and Münte, T.F., 2019. Machine learning based classification of resting-state fMRI features exemplified by metabolic state (hunger/satiety). Frontiers in human neuroscience, 13, p.164.

[8]   McClure I. Volkmar FR, Paul R, Rogers SJ, Pelphrey KA, editors. Handbook of autism and pervasive developmental disorders, fourth edition. Hoboken, NJ: John Wiley & Sons; 2014

[9]   Just, Marcel Adam, Vladimir L. Cherkassky, Timothy A. Keller, and Nancy J. Minshew. "Cortical activation and synchronization during sentence comprehension in high-functioning autism: evidence of underconnectivity." Brain 127, no. 8 (2004): 1811-1821.

[10]  Courchesne, Eric, Karen Pierce, Cynthia M. Schumann, Elizabeth Redcay, Joseph A. Buckwalter, Daniel P. Kennedy, and John Morgan. "Mapping early brain development in autism." Neuron 56, no. 2 (2007): 399-413.

[11]  Kleinhans, Natalia M., Todd Richards, Lindsey Sterling, Keith C. Stegbauer, Roderick Mahurin, L. Clark Johnson, Jessica Greenson, Geraldine Dawson, and Elizabeth Aylward. "Abnormal functional connectivity in autism spectrum disorders during face processing." Brain 131, no. 4 (2008): 1000-1012.

[12]  Weng, Shih-Jen, Jillian Lee Wiggins, Scott J. Peltier, Melisa Carrasco, Susan Risi, Catherine Lord, and Christopher S. Monk. "Alterations of resting-state functional connectivity in the default network in adolescents with autism spectrum disorders." Brain research 1313 (2010): 202-214.

[13]  Comon, P., 1994. Independent component analysis, a new concept?. Signal processing, 36(3), pp.287-314.

[14]  Tošić, I. and Frossard, P., 2011. Dictionary learning. IEEE Signal Processing Magazine, 28(2), pp.27-38.

[15]  Stone, J.V., 2004. Independent component analysis: a tutorial introduction.

[16]  Bordier, C., Dojat, M. and de Micheaux, P.L., 2011. Temporal and spatial independent component analysis for fMRI data sets embedded in the AnalyzeFMRI R package. Journal of Statistical Software, 44(9), pp.1-24.

[17]  Zhao, S., Han, J., Lv, J., Jiang, X., Hu, X., Zhao, Y., Ge, B., Guo, L. and Liu, T., 2015. Supervised dictionary learning for inferring concurrent brain networks. IEEE transactions on medical imaging, 34(10), pp.2036-2045.

[18]  Lv, J., Jiang, X., Li, X., Zhu, D., Chen, H., Zhang, T., Zhang, S., Hu, X., Han, J., Huang, H. and Zhang, J., 2015. Sparse representation of whole-brain fMRI signals for identification of functional networks. Medical image analysis, 20(1), pp.112-134.

[19]  Mairal, J., Bach, F., Ponce, J. and Sapiro, G., 2010. Online learning for matrix factorization and sparse coding. Journal of Machine Learning Research, 11(1).

[20]  Varoquaux, G., Sadaghiani, S., Pinel, P., Kleinschmidt, A., Poline, J.B. and Thirion, B., 2010. A group model for stable multi-subject ICA on fMRI datasets. Neuroimage, 51(1), pp.288-299.

[21]  Mensch, A., Varoquaux, G. and Thirion, B., 2016, April. Compressed online dictionary learning for fast resting-state fMRI decomposition. In 2016 IEEE 13th International Symposium on Biomedical Imaging (ISBI) (pp. 1282-1285). IEEE.

[22]  Calhoun, V.D., Adali, T., Pearlson, G.D. and Pekar, J.J., 2001. A method for making group inferences from functional MRI data using independent component analysis. Human brain mapping, 14(3), pp.140-151.

[23]  Beckmann, C.F., DeLuca, M., Devlin, J.T. and Smith, S.M., 2005. Investigations into resting-state connectivity using independent component analysis. Philosophical Transactions of the Royal Society B: Biological Sciences, 360(1457), pp.1001-1013.